\documentclass[12pt]{article}
\input{amssym}

\def\beq{\begin{equation}}
\def\eeq{\end{equation}}
\def\req#1{(\ref{#1})}
\def\ds{\displaystyle}
\def\text#1{\mbox{\scriptsize{#1}}}
\def\wt#1{\widetilde{#1}}
\def\B{\mbox{B}}
\def\ni{\noindent}

\begin{document}
\baselineskip 0.7cm

\ni{\bf\Large Evaluation of the Casimir Force for a
Dielectric-diamagnetic Cylinder with Light Velocity Conservation
Condition and the Analogue of Sellmeir's Dispersion Law}

\vspace*{1cm}


\ni Iver Brevik\footnote{Department of Energy and Process
Engineering, Norwegian University of Science and Technology,
N-7491 Trondheim, Norway. E-address:  iver.h.brevik@ntnu.no}
 and August Romeo\footnote{Societat Catalana de F{\'\i}sica, Barcelona, Catalonia, Spain}

\vspace*{0.5cm}
\ni PACS numbers: 03.70.+k, 12.20.-m

\vspace{0.5cm}


\vspace*{1cm}

\ni{\bf Abstract.}
We study the Casimir pressure for
a dielectric-diamagnetic cylinder subject to
light velocity conservation and with a dispersion law analogous to
Sellmeir's rule. Similarities to and differences from the spherical case
are pointed out.

\section{Introduction}
The Casimir effect is the force between macroscopic bodies,
conductors or dielectrics, caused by quantum fluctuations in the
vacuum energy of the electromagnetic field or, in a wider sense,
of any field. Particularly, the role of dispersion in vacuum
energies for dielectrics has received considerable attention for
quite a long time (see e.g. Refs.\cite{SDRM}-\cite{cavero06}). In
the present work, the object to be studied  is an infinitely long
cylinder of radius $a$, along the $z$ axis, subject to the
following limitations:
\begin{itemize}
\item The cylinder itself is dielectric-diamagnetic and
dispersive. Its permittivity $\varepsilon$ and permeability $\mu$
will depend on the frequency $\omega$ of each electromagnetic
wave. Light velocity conservation on both sides of the boundary is
imposed, i.e., the cylindrical surface is an interface between two
relativistic media. The physical significance of this condition in
connexion with QCD vacuum was explained in Refs.\cite{L} and
\cite{BE}-\cite{B}. Moreover, this constraint is convenient
because it reduces the number of required mathematical
ingredients.  In particular, there is no need of calculating a
contact term; the Casimir force is derivable from the Maxwell
stress tensor directly.  Using the unit system in which
$\hbar=c=1$, we have \beq \varepsilon(\omega)\mu(\omega)= 1.
\label{samec} \eeq As $\mu(\omega)$ we will choose the analogue of
Sellmeir's rule with one single absorption frequency, given by
Eq.\req{anSell} below. \item The environment of the cylinder is
vacuum, with permittivity and permeability equal to one.
\end{itemize}

Before embarking on the mathematical formalism, we think that it
will be useful to provide some essentials of the background for
this kind of theory.

\vspace{0.4cm}

1)  The theory is constructed in terms of macroscopic parameters
such as permeability and permittivity. These concepts are of
course classical concepts first of all, and it is not evident
beforehand that they will be useful also in a quantum mechanical
formulation of electrodynamics in media. One has first to
calculate the consequences of the theory. What is clear from the
outset, is that the continuum picture of the medium has to break
down at very high frequencies.

2)  When calculating the surface force density on the cylinder,
one encounters a weak divergence when summing over the angular
momentum quantum numbers $m$. We emphasize that this kind of
divergence goes beyond the usual high-energy divergences which can
be cured by using one of the regularization schemes: cutoff
regularization, dimensional regularization, or the  zeta function
regularization (for a treatise on the last method  cf.
Ref.~\cite{elizalde94}). As far as the frequency dependence is
concerned, we handle it in terms of
 a frequency cutoff, either in the form of Sellmeir's
dispersion relation (equation (\ref{anSell}) below), or by an
abruptly cut off  dispersion relation (equation (\ref{x}) below).
Both these dispersion relations are allowed from thermodynamics,
which states that the permittivity/permeability - like any other
generalized susceptibility - has to obey a monotonic decrease with
respect to the
 imaginary frequency $\hat\omega$,  from  $\hat{\omega}=0$ to
 $\hat{\omega}=\infty$
\cite{landau84}. The high energy cutoff implies that there is no
divergence associated with the integration over frequencies from
zero  to infinity.

3) Problems of this type are not new; they are encountered in
nuclear physics also. In connection with the Casimir effect in
media they seem first to have been treated by Candelas
\cite{candelas82}. We emphasized  this kind of behaviour ourselves
also, in Ref.~\cite{BN}. One can deal with the divergence
formally, in an approximate way, by truncating the $m$ sum at an
upper limit $m=m_0$, of the order of $a\omega_0$ where $\omega_0$
is the high-frequency cutoff. There are obvious physical reasons
for truncating the $m$ sum in this way: the cutoff parameter
$\omega_0$ means that the presence of the medium cannot be felt by
photons whose frequencies are much higher than $\omega_0$. A
photon of limiting frequency $\omega_0$ propagating in the
$xy$-plane and touching the cylindrical surface, has an angular
momentum equal to $a\omega_0$. Angular momentum values $m \gg
a\omega_0$ are therefore not expected to be significant.

4)  It turns out that the need of introducing an extra angular
momentum cutoff $m_0$ besides the high-frequency cutoff $\omega_0$
is highly dependent  on the geometry. One has to introduce a
similar angular momentum cutoff in the case of a spherical shell
also, when the material satisfies the same condition
$\varepsilon\mu=1 $ \cite{brevik90}. However, for a {\it compact
spherical ball}, it turns out  that no angular momentum cutoff is
needed \cite{BE,BC}.

5) The appearance of the angular momentum cutoff as an
order-of-magnitude entity may appear surprising, for the following
reason: one might expect that the theoretical surface force on a
cylinder is a clear-cut quantity, as it should be accessible  to
measurement, at least in principle. In our opinion this lack of
exactness reflects that parameters like permittivity and
permeability are lump parameters that are being used to the limit
of their applicability. Ideally speaking, one would desire the
lump-parameter theory to  be replaced by a microscopic many-body
theory, but such an approach would be complicated and most likely
be very little useful in practice.

It should also be observed that there is one physical ingredient
that can be hidden in the macroscopical formalism, namely {\it
surface tension}. Actually, the correspondence between a formal
divergence in the force  and a surface tension, was discussed by
Milton a long time ago, in connection with Casimir theory of a
nondispersive dielectric ball \cite{milton80}. In the calculation
of the surface force there occurred a divergent term proportional
to $\delta^{-3}$, where $\delta=\tau/a$ is a nondimensional
parameter associated with the time splitting $\tau=t-t'$ between
the two space-time points in the photon Green function. See also
the more recent discussions in Ref.~\cite{Mb,cavero06} and related
works, including \cite{RS}.

6) We finally point out that our calculations will apply to the
case of zero temperature only.

\vspace{0.5cm}

 In Sec. 2, several
expressions for the surface force density are derived. After
applying one of them to the dispersion rule in question,
analytical and numerical aspects are discussed in Sec. 3.
Observations about the obtained results have been included in Sec.
4. An appendix contains some details on the nondispersive limit.

\section{Surface force density}
The (lateral) surface force density, or Casimir radial pressure,
was obtained from Maxwell's stress tensor by Brevik and Nyland in
ref.\cite{BN} (for background material about Casimir problems with
cylindrical symmetry, see also Ref.\cite{Mb}). Thus, our starting
point will be formula (69) in Ref.\cite{BN}, which supplies the
surface force density $F$ for arbitrary $\mu(\omega)$: \beq F=
\frac{1}{2 \, \pi^3 \, a^4} \int_{0}^{\infty} dx \int_{0}^{\infty}
dy {\sum_{m=0}^{\infty}}' (\mu-1)\left(
\frac{1}{P_m}+\frac{1}{\wt{P}_m} \right) \, r \, \left[ \left(
1+\frac{m^2}{r^2} \right) I_mK_m+I_m'K_m' \right] . \label{69}
\eeq This formula follows from the solution of Maxwell's equations
subject to the electromagnetic boundary conditions at the surface
$r=a$, using Schwinger's source theory and taking advantage of the
Green function dyadic method. There is no regularization procedure
involved at this stage.

The prime on the summation sign in (\ref{69}) indicates that the
$m=0$ term is counted with half weight, i.e.,
$\ds{\sum_{m=0}^{\infty}}'= \frac{1}{2}\sum_{m=-\infty}^{\infty}$.
Observe that, since $\mu$ depends on the frequency, the $\mu-1$
factor should be inside the integral, not outside. Here \beq
r=\sqrt{x^2+y^2}, \qquad x= -i a \omega, \qquad y= a k, \eeq where
$\omega$ denotes the energy eigenfrequency and $k$ the momentum
along the cylinder axis.  Note that $r$ is not the radial
coordinate, but the momentum component perpendicular to the
cylinder axis, after complex rotation and in nondimensional form.
Similarly, we can define $\varphi= \arctan(y/x) \Longrightarrow x=
r \cos\varphi$ (of course, this $\varphi$ is not the angular
coordinate in ordinary space, but an angle in momentum space).
Doing so, the integrals can be reexpressed as \beq
\int_{0}^{\infty} dx \int_{0}^{\infty} dy \dots =
\int_{0}^{\infty} dr \, r \, \int_{0}^{\pi/2} d\varphi \dots
\label{vc} \eeq $F$ involves the following quantities from
Ref.\cite{BN} \beq
\begin{array}{lll}
P_m&=&I_m'K_m-\mu I_m K_m' \\
\wt{P}_m&=&I_m K_m'-\mu I_m'K_m ,
\end{array}
\eeq
where all the unwritten arguments are $r$'s.
Bearing in mind the fact that $I_m$, $K_m$ are solutions of the
modified Bessel equation, one verifies that
\beq
\begin{array}{lll}
\ds\left( \frac{\partial P_m}{\partial r} \right)_{\mu}&=&\ds -\frac{1}{r}P_m
+(1-\mu)\left[
\left( 1+\frac{m^2}{r^2} \right) I_mK_m+I_m'K_m' \right] \\
\ds\left( \frac{\partial \wt{P}_m}{\partial r} \right)_{\mu}&=&\ds -\frac{1}{r}\wt{P}_m
+(1-\mu)\left[
\left( 1+\frac{m^2}{r^2} \right) I_mK_m+I_m'K_m' \right] .
\end{array}
\label{dPms}
\eeq
The notation $\left( \frac{\partial f}{\partial r} \right)_{\mu}$
indicates differentiation of $f$ with respect to $r$ keeping the $\mu$
function as if it were independent of $r$.
Using \req{dPms}, we find
\beq
\left( \frac{\partial}{\partial r} \ln( r^2 \, P_m \, \wt{P}_m ) \right)_{\mu}=
(1-\mu)\left( \frac{1}{P_m}+\frac{1}{\wt{P}_m} \right)
\left[ \left( 1+\frac{m^2}{r^2} \right) I_mK_m+I_m'K_m' \right] .
\label{mfi}
\eeq
The r.h.s. happens to be the main factor in the integrand of \req{69}.
As a result of \req{vc}, \req{mfi} we see that the $F$ given by \req{69}
may be recast into the form
\beq
F= -\frac{1}{2 \, \pi^3 \, a^4}
\int_{0}^{\infty} dr
\int_{0}^{\pi/2} d\varphi \,
r^2 \, {\sum_{m=0}^{\infty}}' \,
\left( \frac{\partial}{\partial r} \ln( r^2 \, P_m \, \wt{P}_m ) \right)_{\mu}.
\label{69b}
\eeq
Further, in terms of the new notation
\beq
\begin{array}{lll}
\mathcal{W}_m^2&=&\ds(I_mK_m'-I_m'K_m)^2=\frac{1}{r^2} \\
\mathcal{P}_m^2&=&((I_mK_m)')^2 ,
\end{array}
\label{defsWP}
\eeq
($\mathcal{P}_m$ should not be mistaken for $P_m$)
one realizes that
\beq
\begin{array}{lll}
r^2 \, P_m \, \wt{P}_m&=&\ds\frac{r^2}{4}\left[
(1-\mu)^2 \mathcal{P}_m^2 -(1+\mu)^2 \mathcal{W}_m^2
\right] \\
&=&\ds-\frac{(1+\mu)^2}{4} ( 1-\xi^2 r^2 \mathcal{P}_m^2 ) ,
\end{array}
\label{r2PmPm}
\eeq
where we have introduced
\beq
\xi^2\equiv \left(
\frac{\varepsilon_1-\varepsilon_2}{\varepsilon_1+\varepsilon_2}
\right)^2
= \left( \frac{1-\mu}{1+\mu} \right)^2.
\label{defxi2}
\eeq
$\xi^2$ is, in general, a function of $\omega$ through $\mu(\omega)$.
By virtue of \req{r2PmPm},
\beq
\begin{array}{lll}
\ds\left( \frac{\partial}{\partial r} \ln( r^2 \, P_m \, \wt{P}_m ) \right)_{\mu}
&=&\ds\left( \frac{\partial}{\partial r}
\ln\left[  1 - \xi^2 r^2 \mathcal{P}_m^2 \right]
\right)_{\mu} .
\label{ddrln}
\end{array}
\eeq Inserting \req{ddrln} into \req{69b} and replacing
$\mathcal{P}_m^2$ with its explicit expression \beq F= -\frac{1}{2
\, \pi^3 \, a^4} \int_{0}^{\infty} dr \int_{0}^{\pi/2} d\varphi \,
r^2 \,{\sum_{m=0}^{\infty}}' \,  \left( \frac{\partial}{\partial
r} \ln\left[ 1 - \xi^2 r^2 ((I_m(r)K_m(r))')^2 \right]
\right)_{\mu}. \label{69c} \eeq As commented, $\mu^2$, $\xi^2$ are
in general functions of $\omega= i a^{-1} x = i a^{-1} r
\cos\varphi$. In the nondispersive case they are just constants,
and the $\varphi$ integration is trivially factored out, yielding
$\ds\int_{0}^{\pi/2} d\varphi = \pi/2$. Thus, after recalling the
meaning of $\ds{\sum_m}'$, $F$ reduces to \beq F= -\frac{1}{8 \,
\pi^2 \, a^4} \int_{0}^{\infty} dr \, r^2 \,
\sum_{m=-\infty}^{\infty} \, \frac{d}{dr} \ln\left[  1 - \xi^2 r^2
((I_m(r)K_m(r))')^2 \right] . \eeq Since, according to
Ref.\cite{BN}, this result coincides with the 'radial pressure'
$P$ given by formula 67 in Ref.\cite{CPM} (which, in turn, is a
generalization of the expression found in Ref.\cite{RaMi}).

It may be interesting to rewrite \req{69c} in the same fashion as
Brevik and Einevoll did in Ref.\cite{BE} for the case of the ball.
Differentiating with fixed $\mu$ and using \req{r2PmPm}, one finds
\beq \left( \frac{\partial}{\partial r} \ln\left[ 1 - \xi^2 r^2
\mathcal{P}_m^2 \right] \right)_{\mu} =\frac{1}{2} \chi^2 \frac{
\ds\frac{d}{dr}(r^2\mathcal{P}_m^2) }{ r^2 \, P_m \, \wt{P}_m } ,
\label{BE1} \eeq where \beq \chi\equiv \chi(\omega)=
\mu(\omega)-1. \eeq On the other hand, \beq \left(
\frac{\partial}{\partial r} \ln\left[ 1 - \xi^2 r^2
\mathcal{P}_m^2 \right] \right)_{\mu}= \xi^2 \frac{1-r^2 \,
\mathcal{P}_m^2}{1-\xi^2 \, r^2 \, \mathcal{P}_m^2}
\frac{d}{dr}\ln( 1-r^2 \mathcal{P}_m^2 ) . \label{prefq} \eeq
Next, we observe that \beq \frac{1}{\xi^2}\frac{1-\xi^2 \, r^2 \,
\mathcal{P}_m^2} {1-r^2 \, \mathcal{P}_m^2} = \left(
1-\ds\frac{\chi^{-1}}{r I_m K_m'} \right) \left(
1+\ds\frac{\chi^{-1}}{r I_m' K_m} \right) . \label{fq} \eeq All
the time, the unwritten arguments of $\mathcal{P}_m$, $I_m$,
$K_m$, $I_m'$, $K_m'$ are $r$'s. Inserting \req{BE1} and the
combination of \req{prefq},\req{fq} into \req{69c}, one obtains
\beq
\begin{array}{lll}
F&=&\ds -\frac{1}{4 \, \pi^3 \, a^4}
\int_{0}^{\pi/2} d\varphi \,
\int_{0}^{\infty} dr \, r^2 \, \chi^2 \, {\sum_{m=0}^{\infty}}'
\frac{ \ds\frac{d}{dr}(r^2\mathcal{P}_m^2) }{ r^2 \, P_m \, \wt{P}_m } \\
&=&\ds -\frac{1}{2 \, \pi^3 \, a^4}
\int_{0}^{\pi/2} d\varphi \,
\int_{0}^{\infty} dr \, r^2 \,{\sum_{m=0}^{\infty}}'
\frac{ \ds\frac{d}{dr}\ln( 1-r^2 \mathcal{P}_m^2 ) }
{\left( 1-\ds\frac{\chi^{-1}}{r I_m K_m'} \right)
\left( 1+\ds\frac{\chi^{-1}}{r I_m' K_m} \right)} ,
\end{array}
\label{an311} \eeq which are analogous to formulas (3.11a) and
(3.11b) in Ref.\cite{BE}, but contain a new element: the $\varphi$
dependence ---in addition to the $r$ dependence--- of the $\chi$
function, since $\chi(\omega)=\chi(i a^{-1} r\cos\varphi )$. In
particular, choosing the $\mu(\omega)$ analogous to Sellmeir's
$\varepsilon(\omega)$ in ordinary electromagnetism, \beq
\chi(\omega)=\mu(\omega)-1=\frac{\mu_0-1}{1-\omega^2/\omega_0^2}
\label{anSell} \eeq we are led to \beq
\chi=\frac{\chi_0}{1+\ds\frac{r^2}{x_0^2}\cos^2\varphi}, \qquad
\chi_0=\mu_0-1, \qquad x_0= a\omega_0 . \eeq Taking again
\req{69c}, \req{defsWP} and relation \req{prefq}, one can arrive
at \beq F= -\frac{1}{2 \, \pi^3 \, a^4} {\sum_{m=0}^{\infty}}'
\int_{0}^{\pi/2} d\varphi \, \int_{0}^{\infty} dr \, r^2 \, \xi^2
\left[ 1+(\xi^2-1)\frac{r^2 \mathcal{P}_m^2}{1-\xi^2
r^2\mathcal{P}_m^2} \right] \frac{d}{dr} \ln( 1-r^2
\mathcal{P}_m^2 ), \label{Fnew} \eeq where, according to the
definition \req{defxi2} and our choice \req{anSell} \beq
\begin{array}{c}
\ds\xi^2\equiv\xi^2(r,\varphi)= \xi_0^2
\left( \frac{1}{1+\alpha^{-2}(r) \, \cos^2\varphi} \right)^2, \\
\ds\xi_0^2\equiv \left( \frac{\mu_0-1}{\mu_0+1} \right)^2, \qquad
\alpha(r)\equiv \left( \frac{\mu_0+1}{2} \right)^{1/2} \frac{x_0}{r} .
\end{array}
\label{xi2rphi}
\eeq

\section{Calculation and results}
\subsection{Uniform asymptotic expansions for $m\neq 0$\label{ssuae}}
Before proceeding to any numerical evaluation,
it is convenient to study the behaviour of \req{Fnew}
by means of (Debye) uniform asymptotic expansions \cite{AS}.
Leaving the $m=0$ contribution aside, for $m\neq 0$ we may rescale
$r\to mr$ in the integrand of \req{Fnew}
and obtain its Debye expansion for $mr \gg 1$. Starting from
\beq
\begin{array}{rcl}
\ds r^2 \mathcal{P}_m^2|_{r\to mr}&\sim&\ds U_2(t(r))\frac{1}{m^2}
+{\cal O}\left( \frac{1}{m^4} \right) \\
U_2(t)&=&\ds\frac{t^2}{4}-\frac{t^4}{2}+\frac{t^6}{4} \\
t(r)&\equiv&\ds\frac{1}{\sqrt{1+r^2} },
\label{defU2}
\end{array}
\eeq one gets \beq \left. 1+(\xi^2-1)\frac{r^2
\mathcal{P}_m^2}{1-\xi^2 r^2\mathcal{P}_m^2} \right\vert_{r\to
mr}\sim 1+(\xi^2(mr,\varphi)-1) \left[ U_2(t(r))\frac{1}{m^2}
+{\cal O}\left( \frac{1}{m^4} \right) \right] \eeq ($U_n$, $V_n$
should not be mistaken for the $u_n$, $v_n$ of Ref.\cite{AS},
although the former are obtained from the latter). Further, \beq
\begin{array}{rcl}
\ds\left. \frac{d}{dr} \ln( 1-r^2 \mathcal{P}_m^2 ) \right\vert_{r\to mr}&\sim&\ds
\frac{r}{m} \left[ V_2(t(r))\frac{1}{m^2}+V_4(t(r))\frac{1}{m^4}
+{\cal O}\left( \frac{1}{m^6} \right) \right] \\
V_2(t)&=&\ds\frac{t^4}{2}-2 t^6+\frac{3}{2} t^8  \\
V_4(t)&=&\ds\frac{7}{8}t^6 -\frac{57}{4}t^8 +\frac{101}{2} t^{10}
-\frac{252}{4} t^{12} +\frac{213}{8} t^{14} .
\end{array}
\label{defVs}
\eeq
Hence, we find
\beq
\begin{array}{c}
\ds F_{ \{ m\neq 0 \} }\sim -\frac{1}{2 \, \pi^3 \, a^4}
\sum_{m=1}^{\infty} m^2 \,
\int_{0}^{\pi/2} d\varphi \,
\int_{0}^{\infty} dr \, r^3 \, \xi^2(mr, \varphi) \\
\ds\times\left\{
V_2(t) \frac{1}{m^2}
+ \left[ V_4(t)+( \xi^2(mr, \varphi)-1 ) \,
U_2(t) \, V_2(t) \right]
\frac{1}{m^4}
+{\cal O}\left( \frac{1}{m^6} \right)
\right\} ,
\end{array}
\eeq where $t \equiv t(r)$ should be understood. Adopting a
notation reminiscent of Ref.\cite{BE}, this is rewritten as \beq
\begin{array}{rcl}
\ds F_{ \{ m\neq 0 \} }&\sim&\ds -\frac{1}{2 \, \pi^3 \, a^4}
\sum_{m=1}^{\infty}
\left[
\mathcal{J}_m^{(0)}+\mathcal{J}_m^{(2)}\frac{1}{m^2}
+\mathcal{I}_m\frac{1}{m^2}+{\cal O}\left( \frac{1}{m^4} \right)
\right] \\
\mathcal{J}_m^{(0)}&=&\ds \int_{0}^{\pi/2} d\varphi \,
\int_{0}^{\infty} dr \, r^3 \, \xi^2(mr, \varphi) \, V_2(t) \\
\mathcal{J}_m^{(2)}&=&\ds \int_{0}^{\pi/2} d\varphi \,
\int_{0}^{\infty} dr \, r^3 \, \xi^2(mr, \varphi) \, V_4(t) \\
\mathcal{I}_m&=&\ds \int_{0}^{\pi/2} d\varphi \,
\int_{0}^{\infty} dr \, r^3 \, \xi^2(mr, \varphi) \,
(\xi^2(mr, \varphi)-1) \, U_2(t) \, V_2(t) .
\end{array}
\label{defJmsIm}
\eeq
Let's consider the $\varphi$ integrations.
Recalling \req{xi2rphi} for $r\to mr$,
\beq
\begin{array}{lll}
\mathcal{J}_m^{(0)}&=&\ds
\xi_0^2
\int_{0}^{\infty} dr \, r^3 \, V_2(t)
\int_{0}^{\pi/2} d\varphi \,
\frac{1}{ ( 1+m^2\alpha^{-2}\cos^2\varphi )^2 } \\
\mathcal{J}_m^{(2)}&=&\ds
\xi_0^2
\int_{0}^{\infty} dr \, r^3 \, V_4(t)
\int_{0}^{\pi/2} d\varphi \,
\frac{1}{ ( 1+m^2\alpha^{-2}\cos^2\varphi )^2 } \\
\mathcal{I}_m&=&\ds
\int_{0}^{\infty} dr \, r^3 \, U_2(t) \, V_2(t)
\left[
\xi_0^4
\int_{0}^{\pi/2} d\varphi \,
\frac{1}{ ( 1+m^2\alpha^{-2}\cos^2\varphi )^4 } \right. \\
&&\hspace{9em}\ds\left. -\xi_0^2
\int_{0}^{\pi/2} d\varphi \,
\frac{1}{ ( 1+m^2\alpha^{-2}\cos^2\varphi )^2 }
\right] ,
\end{array}
\label{JsI}
\eeq
where $\alpha$ stands for the $\alpha(r)$ defined in \req{xi2rphi}.
The values of the occurring $\varphi$ integrals are
\beq
\begin{array}{rcl}
\ds\int_{0}^{\pi/2} d\varphi \,
\frac{1}{( 1+m^2\alpha^{-2}\cos^2\varphi )^2 }&=&
\ds\frac{\pi}{4} \frac{\alpha ( m^2+2\alpha^2 )}{(m^2+\alpha^2)^{3/2}} \\
\ds\int_{0}^{\pi/2} d\varphi \,
\frac{1}{( 1+m^2\alpha^{-2}\cos^2\varphi )^4 }&=&
\ds\frac{\pi}{16}  \frac{\alpha ( m^2+2\alpha^2 ) (5m^4+8\alpha^2m^2+8\alpha^4)}
{(m^2+\alpha^2)^{7/2} } .
\end{array}
\label{intphis} \eeq At this stage one may wonder about the
possibility of performing the $m$ summation. For
$\sum_m\mathcal{J}_m^{(0)}$, simple power counting shows that the
 harmonic series is divergent. Such an obstacle will be avoided,
 as discussed in the Introduction,
by truncating the $m$ summation as was done in  Ref.\cite{BN}.
Thus, the only contributing modes will be those for which  $m \le
m_0$, where \beq m_0= f x_0, \label{defm0} \eeq  $f$ being a
factor of order unity. Physically, macroscopic electromagnetic
theory seems to be too crude to yield a definite theoretical
expression for the surface force.

It should be noted that the above points are not dependent on the
particular form  (\ref{anSell}) that we have chosen for the
dispersion relation. This relation implies a soft high frequency
cutoff. The same kind of behaviour is found if we assume instead
an abrupt cutoff, in the form of a dispersion equation reading
\begin{equation}
\mu(i\hat{\omega})=\left\{ \begin{array}{ll} \mbox{constant}, &
\hat{\omega} \leq \omega_0 \\
1                                                          , &
\hat{\omega}
> \omega_0,
\end{array}
\right. \label{x}
\end{equation}
$\hat{\omega}$ being the imaginary frequency. As discussed above,
this dispersion relation means that the medium cannot be felt by
photons whose frequencies are larger than $\omega_0$.

Now, instead of summing over  $m$, we substitute \req{intphis}
into \req{JsI}, perform the $r$ integration and obtain the
following results \beq
\begin{array}{lll}
\mathcal{J}_m^{(0)}&=&\ds\frac{\pi}{4}\xi_0^2
\sum_{q=2}^{4} V_{2,q} \left[
2 \, I_4\left( 3, \frac{3}{2}, q , \alpha_m^{-2} \right)
+\alpha_m^{-2} \, I_4\left( 5, \frac{3}{2}, q , \alpha_m^{-2} \right)
\right]  \\
\mathcal{J}_m^{(2)}&=&\ds\frac{\pi}{4}\xi_0^2
\sum_{q=3}^{7} V_{4,q} \left[
2 \, I_4\left( 3, \frac{3}{2}, q , \alpha_m^{-2} \right)
+\alpha_m^{-2} \, I_4\left( 5, \frac{3}{2}, q , \alpha_m^{-2} \right)
\right]  \\
\mathcal{I}_m&=&\ds\sum_{q=3}^{7} (U_2V_2)_{,q} \left\{
\frac{\pi}{32}\xi_0^4 \left[
16 \, I_4\left( 3, \frac{7}{2}, q , \alpha_m^{-2} \right)
+24\alpha_m^{-2} \, I_4\left( 5, \frac{7}{2}, q , \alpha_m^{-2} \right)
\right. \right. \\
&&\ds\hspace{8em}\left. +18\alpha_m^{-4} \, I_4\left( 7, \frac{7}{2}, q, \alpha_m^{-2} \right)
+5\alpha_m^{-6} \, I_4\left( 9, \frac{7}{2}, q , \alpha_m^{-2} \right)
\right] \\
&&\ds\hspace{5em}\left. -\frac{\pi}{4}\xi_0^2
\left[
2 \, I_4\left( 3, \frac{3}{2}, q , \alpha_m^{-2} \right)
+\alpha_m^{-2} \, I_4\left( 5, \frac{3}{2}, q , \alpha_m^{-2} \right)
\right] \right\} ,
\end{array}
\label{JmsIm}
\eeq
where
\beq
\alpha_m^{-2}\equiv \alpha^{-2}(m)= \frac{2}{\mu_0+1}\frac{m^2}{x_0^2},
\label{defalm}
\eeq
$V_{2,q}$, $V_{4,q}$, $(U_2V_2)_{,q}$
denote the coefficients of $t^{2q}$ in the polynomials
$V_2(t)$, $V_4(t)$,
$U_2(t) V_2(t)$, respectively (see \req{defU2},\req{defVs}),
and
\beq
\begin{array}{c}
\ds I_4(A,B,q, \beta)\equiv
\int_{0}^{\infty} dr \, r^{A} \, (1+\beta r^2)^{-B} \, (1+r^2)^{-q} \\
\ds =\frac{1}{2} \beta^{-B} \,
\B\left( -\frac{A}{2}+B+q+\frac{1}{2}, \frac{A+1}{2} \right)
{}_2F_1\left( -\frac{A}{2}+B+q+\frac{1}{2}, B; B+q; 1-\frac{1}{\beta} \right).
\end{array}
\eeq
$\B$ and ${}_2F_1$ stand for the Euler beta function and the
hypergeometric function.

\subsection{Asymptotic expansion for $m=0$ \label{ssae}}
In principle, we could content ourselves with a purely numerical evaluation
of the $m=0$ term and, for practical purposes, that would suffice.
At the same time, for coherence, one might like to obtain something
similar to what has been found for $m\neq 0$.
Indeed, when $m=0$, the integrand in \req{Fnew} can be approximated by
\beq
r^3 \xi^2(r,\varphi) \left[ \frac{1}{2r^4}
+\left( \frac{7}{8} +( \xi^2(r,\varphi)-1)\frac{1}{8}\right) \frac{1}{r^6}
+\mathcal{O}\left( \frac{1}{r^8} \right) \right]
\label{intg0as}
\eeq
for large values of $r$. Of course, this replacement cannot be made
near $r=0$. Among the possible ways of dealing with this
issue, we choose to split up the $r$ integration
range into $(0,r_a)$ and $(r_a, \infty)$, using the asymptotic form
\req{intg0as} only in the second domain, while the first part is
numerically evaluated. Fairly good results are obtained for
$r_a= 3$. The outcome may be expressed as follows
\beq
F_{m=0}\sim -\frac{1}{2 \, \pi^3 \, a^4}
\frac{1}{2}\left[
\int_{0}^{\pi/2} d\varphi \,
\int_{0}^{r_a} dr \, \mbox{exact integrand $(m=0)$}
+J_0^{(0)}+J_0^{(2)}+I_0
\right] ,
\eeq
where
\beq
\begin{array}{lll}
J_0^{(0)}&=&\ds\frac{1}{2}\int_{0}^{\pi/2} d\varphi \,
\int_{r_a}^{\infty} dr  \, r^{-1}
\xi^2(r, \varphi), \\
J_0^{(2)}&=&\ds\frac{7}{8}\int_{0}^{\pi/2} d\varphi \,
\int_{r_a}^{\infty} dr \, r^{-3}
\xi^2(r, \varphi), \\
I_0&=&\ds\frac{1}{8}\int_{0}^{\pi/2} d\varphi \,
\int_{r_a}^{\infty} dr \, r^{-3}
\xi^2(r, \varphi)(\xi^2(r, \varphi)-1) , \\
\end{array}
\label{defJ0sI0}
\eeq
which, after performing the integrations, become
\beq
\begin{array}{lll}
J_0^{(0)}&=&\ds\frac{\pi}{8}\xi_0^2
\left[
2 \, I_3\left( -1, \frac{3}{2}, \frac{r_a}{\alpha_1}  \right)
+I_3\left( 1, \frac{3}{2}, \frac{r_a}{\alpha_1} \right)
\right]  \\
J_0^{(2)}&=&\ds\frac{7\pi}{32}\xi_0^2 \, \alpha_1^{-2} \,
\left[
2 \, I_3\left( -3, \frac{3}{2}, \frac{r_a}{\alpha_1}  \right)
+I_3\left( -1, \frac{3}{2}, \frac{r_a}{\alpha_1} \right)
\right]  \\
I_0&=&\ds\frac{\pi}{256}\xi_0^4 \, \alpha_1^{-2} \,  \left[
16 \, I_3\left( -3, \frac{7}{2}, \frac{r_a}{\alpha_1} \right)
+24 \, I_4\left( -1, \frac{7}{2}, \frac{r_a}{\alpha_1}  \right)
\right. \\
&&\ds\hspace{5em}\left. +18 \, I_3\left( 1, \frac{7}{2}, \frac{r_a}{\alpha_1} \right)
+5 \, I_3\left( 3, \frac{7}{2}, \frac{r_a}{\alpha_1} \right)
\right] \\
&&\ds -\frac{\pi}{32}\xi_0^2 \, \alpha_1^{-2} \,
\left[
2 \, I_3\left( -3, \frac{3}{2}, \frac{r_a}{\alpha_1} \right)
+ I_3\left( -1, \frac{3}{2}, \frac{r_a}{\alpha_1}  \right)
\right]
\end{array}
\eeq
(note formal similarity to eqs.\req{JmsIm}).
Here $\alpha_1$ is given by definition \req{defalm} for $m=1$, and
\beq
\begin{array}{lll}
\ds I_3(A,B,R)&\equiv&\ds\int_{R}^{\infty} dr \, r^A \, (1+r^2)^{-B}
=\frac{1}{2} \,
\B_{1/(1+R^2)}\left( -\frac{A}{2}+B-\frac{1}{2}, \frac{1+A}{2} \right) \\
&=&\ds\frac{1}{2} \, \left( -\frac{A}{2}+B-\frac{1}{2} \right)^{-1}
\left( \frac{1}{1+R^2} \right)^{-A/2+B-1/2}  \\
&&\ds \times {}_2F_1\left( -\frac{A}{2}+B-\frac{1}{2}, \frac{1-A}{2};
-\frac{A}{2}+B+\frac{1}{2}; \frac{1}{1+R^2} \right) ,
\end{array}
\eeq with $\B_x$ denoting the incomplete Beta function of
parameter $x$, which can, in turn, be expressed by means of
another hypergeometric function.

\subsection{Numerical results}
Separate contributions from each $m$ may be numerically obtained
by means of \req{Fnew}-\req{xi2rphi}.
Writing
$F={\sum_{m=0}^{m_0}}' F_m$, we consider
$F_{m \, \text{rel}}=F_m/F_{\text{DRM}}$,
$F_{\text{rel}}=F/F_{\text{DRM}}$,
being
$F_{\text{DRM}}= -0.01356/(\pi a^4)$ the surface force density for the
perfectly conducting case \cite{RaMi}, here taken as reference value.
In table \ref{t51} we list
results for $x_0=5$, $f=1$, and several $\mu_0$ options.
Further, the relative surface force density found by using the
approximations of subsecs. \ref{ssuae}, \ref{ssae} is included as
$F_{\text{rel}}^{\text{app}}$.
\begin{table}[htb]
\begin{center}
\begin{tabular}{|l|r|r|r|r|}
\hline\hline
$m$&\multicolumn{4}{c|}{$F_{m \, \text{rel}}$} \\
&$\mu_0=0.2$&$\mu_0=0.4$&$\mu_0=0.6$&$\mu_0=0.8$ \\ \hline
0& 1.519 & 0.624 & 0.215 & 0.043 \\
1& 0.792 & 0.351 & 0.127 & 0.026 \\
2& 0.404 & 0.182 & 0.066 & 0.014 \\
3& 0.261 & 0.117 & 0.043 & 0.009 \\
4& 0.191 & 0.086 & 0.031 & 0.007 \\
5& 0.150 & 0.067 & 0.025 & 0.005 \\ \hline
$F_{\text{rel}}$& 3.317 & 1.427 & 0.507 & 0.104 \\
$F_{\text{rel}}^{\text{app}}$& 3.298 & 1.420 & 0.505 & 0.104 \\
\hline\hline
\end{tabular}
\end{center}
\caption{Values of $F_{m \, \text{rel}}=F_m/F_{\text{DRM}}$,
$F_{\text{rel}}={\sum_{m=0}^{m_0}}' F_{m \, \text{rel}}$
and $F_{\text{rel}}^{\text{app}}$
(approximation to $F_{\text{rel}}$ using formulas in
subsecs. \ref{ssuae}, \ref{ssae}) for $x_0=5$, $f= 1$, i.e. $m_0=5$,
and $\mu_0= 0.2, 0.4, 0.6, 0.8$. All quantities vanish at $\mu_0=1$.}
\label{t51}
\end{table}
Matching figures for $x_0=50$, $f=1$ are listed in table
\ref{t501}.
\begin{table}[htb]
\begin{center}
\begin{tabular}{|l|r|r|r|r|}
\hline\hline
$m$&\multicolumn{4}{c|}{$F_{m \, \text{rel}}$} \\
&$\mu_0=0.2$&$\mu_0=0.4$&$\mu_0=0.6$&$\mu_0=0.8$ \\ \hline
0& 3.091& 1.267 &0.433& 0.086  \\
1& 3.361& 1.428 &0.498& 0.100  \\
2& 2.519& 1.080 &0.379& 0.077  \\
3& 2.038& 0.880 &0.310& 0.063  \\
4& 1.712& 0.743 &0.263& 0.054  \\
5& 1.472& 0.642 &0.229& 0.047  \\
\vdots&\vdots&\vdots&\vdots&\vdots \\
\hline
$F_{\text{rel}}$& 32.011 & 13.992 & 5.003 & 1.055  \\
$F_{\text{rel}}^{\text{app}}$& 32.064 & 14.017 & 5.012 & 1.034  \\
\hline\hline
\end{tabular}
\end{center}
\caption{First values of $F_{m \, \text{rel}}$, together with
$F_{\text{rel}}$, $F_{\text{rel}}^{\text{app}}$
for $x_0=50$, $f= 1$ ($m_0=50$), $\mu_0= 0.2, 0.4, 0.6, 0.8$.
As can be appreciated, the contributions from $5 < m \le m_0$ are quite
important.}
\label{t501}
\end{table}

\section{Conclusions and Further Discussion}

Let us summarize, and comment further upon,  some of the main
points:
\begin{itemize}

\item The quoted results show that the strength of the force
increases as the value of $x_0$ grows, and that the relative
importance of each $m$ mode diminishes with increasing $m$,
although this decrease is not quick enough for the $m$ series to
converge. Since all the relative figures are positive, and
$F_{\text{DRM}} < 0$, the surface force is {\it negative} for all
the studied cases. Further, this attractiveness persists for each
value of $m$ separately.

\item Unlike the case of the sphere in
Ref.\cite{BE}, the angular momentum summation leads to a
divergence, which we curtail by setting the bound \req{defm0}.
This property could be inherent to the present geometry, as the
same behaviour was already observed, as we have mentioned,  in
Ref.\cite{BN} for a cylinder having a step function  dispersion
relation. If one tries to find the limit as $x_0$ goes to
infinity, the term corresponding to the nondispersive part in
$\mathcal{J}_m^{(0)}$ brings about a divergence as well. However,
known results for nondispersive dielectrics are finite
\cite{MNN,KR}, not divergent, and this finiteness is achieved
without truncating the $m$ summation. This is so because these
calculations introduce regularization procedures which, in fact,
eliminate such infinities (for details, see Appendix \ref{appad}).
Putting it in a another way: the divergences which may be
envisaged as limits of dispersive models when dispersion tends to
zero are removed.

\item One special approach is that of employing the heat kernel
technique. This approach is often useful when discussing
ultraviolet divergences. Bordag et al.\cite{bordag98} considered a
body with permittivity being a smooth function of the coordinates,
and found the decisive heat kernel coefficient $a_2$ to be
nonzero. This corresponds to an infinite Casimir energy, however.
How to obtain a finite answer from this approach is not known.
Again, this behaviour reflects the missing link between
macroscopic electromagnetic theory and experiment.

\item Perhaps can some insight be obtained by considering instead
a {\it scalar} field. For example, we mention that Graham
et.al.\cite{graham04} showed that the Casimir pressure for a
scalar field outside a sphere diverges as the surface of the
sphere becomes sharp, independent of the frequency integration. If
the surface is smoothed out the pressure is finite, but depends on
the shape of
the smoothing function and the distance
scale over which the transition from vacuum to material takes
place. Another work is that of Cavero-Pel\'{a}ez {\it et al.}
\cite{cavero06a}; they considered a massless scalar field
associated with step-function potentials in spherical geometry,
 and found that for zero shell thickness the energy
has a contribution not only from the local energy density but also
from an energy term residing entirely in the surface.

\item In view of the lack of an exact expression for the Casimir
surface force on a single cylinder, it may be natural to look for
corresponding theories for the case of two cylinders. For
instance, Nyland and Brevik \cite{nyland94} calculated the
interaction between two parallel cylinders placed in a surrounding
fluid assumed to possess magnetic as well as dielectric
properties; this work generalizing earlier work of Mitchell {\it
et al.} \cite{mitchell73} dealing with the dielectric case only.
This calculation led to finite results. More recently, the case of
two perfectly conducting concentric cylinders was considered by
Mazzitelli \cite{mazzitelli04}, employing a mode summation method.
The case of eccentric cylinders was considered by Dalvit {\it et
al.} \cite{dalvit06}. The Casimir interaction between a cylinder
and a plate was considered by Emig {\it et al.} \cite{emig06} and
by Bordag \cite{bordag06}, and the interaction between two plates
situated inside a cylinder was recently studied by Marachevsky
\cite{marachevsky07}. Characteristic for all these two-body cases
is that the expressions for the interaction are finite.

\item Thus, when two or more media are involved, the macroscopic
theory works well. The physical reason for the single-medium
problem, as mentioned already in the Introduction, is most likely
that {\it surface tension} is hidden in the formalism. Typically,
the tension is contained in some parameter occurring in the
macroscopic theory, like the time splitting parameter; cf. the
discussions in Refs.~\cite{milton80} and \cite{Mb}. How to relate
the description in terms of lump parameters to a more fundamental
microscopic theory, is at present an open issue.

\item Finally, as a general remark, we note that vacuum energy and
Casimir forces are versatile concepts which have led to a wide
range of implications, not only in pure quantum field theory (e.g.
the subject of flux tubes between quarks in Ref.\cite{KS} and
references  therein), but also in an area like cosmology, and
concerning issues such as the cosmological constant, black holes
or spacetime foam \cite{WG}. Radiation properties in dielectric
cylinders and their influence on accelerator physics have been
recently discussed in Ref.\cite{S}.

\end{itemize}

\appendix\section{Appendix: finiteness of the nondispersive limit
for some analytical regularizations \label{appad}} As already
argued, the possible divergences are originated in the behaviour
of the $\mathcal{J}_m^{(0)}$ objects defined in \req{defJmsIm}
(for $x_0\to\infty$, the $J_0^{(0)}$ from formula \req{defJ0sI0}
will also come into play). Here we shall examine them in the
nondispersive limit, which corresponds to $x_0\to\infty$. Let's
assume that the force is regularized by changing the overall power
of $r$ in the integrand of \req{Fnew} as follows: \beq r^2 \to
r^{1-s}, \label{intrs} \eeq being $s$ a complex parameter for
analytical continuation. This type of power change is what
essentially happens when applying dimensional or zeta-function
regularization (like in, e.g., Ref.\cite{KR}). First, $s$ is
assumed large enough for all expressions to make sense and,
eventually, the analytical extension of the emerging functions to
$s=-1$ is found.

When considering the
$\sum_{m=1}^{\infty} \mathcal{J}_m^{(0)}$ part,
one realizes that,
as a result of the $r \to mr$ rescaling, $m$ powers involving $s$
show up:
\beq
\sum_{m=1}^{\infty} \mathcal{J}_m^{(0)}
= \sum_{m=1}^{\infty} m^{-(s+1)} \,
\int_{0}^{\infty} dr \, r^{2-s} \, V_2(t) \,
\int_{0}^{\pi/2} d\varphi \, \xi^2(mr, \varphi) .
\eeq
Going back to the first integral in \req{intphis}, we take the limit as
$x_0\to\infty$ (i.e., $\alpha\to\infty$) and find
\beq
\int_{0}^{\pi/2} d\varphi \, \xi^2(mr, \varphi) \longrightarrow
\xi_0^2 \, \frac{\pi}{2}.
\label{limiphi}
\eeq
Therefore, reintroducing the explicit form of $t$ as a function $r$
(see \req{defU2}),
\beq
\begin{array}{rrl}
\ds\sum_{m=1}^{\infty} \mathcal{J}_m^{(0)}&\longrightarrow&\ds\xi_0^2 \,
\frac{\pi}{2} \, \sum_{m=1}^{\infty} m^{-(s+1)}
\sum_{q=2}^{4} V_{2,q} \,
\int_{0}^{\infty} dr \, r^{2-s} \, \frac{1}{(1+r^2)^q} \\
&=&\ds\xi_0^2 \, \frac{\pi}{4} \, \zeta_R(s+1) \,
\sum_{q=2}^4 V_{2,q} \, \B\left( \frac{3-s}{2}, \frac{2q-3+s}{2} \right) ,
\end{array}
\eeq
where $V_{2,q}$ are the coefficients of the $V_2$ polynomial given
in \req{defVs} and $\zeta_R$ indicates the Riemann zeta function.
Inspection of the beta function arguments shows that
the only divergent term at $s=-1$ is the one with $q=2$.
Since
$V_{2,2}=1/2$, one has
\beq
\sum_{m=1}^{\infty} \mathcal{J}_m^{(0)}\longrightarrow\xi_0^2 \, \frac{\pi}{8} \,
\left[
\zeta_R(s+1) \, \B\left( \frac{3-s}{2}, \frac{s+1}{2} \right)
+\mathcal{O}((s+1)^0)
\right] ,
\label{limsumJm}
\eeq
where $\mathcal{O}((s+1)^0)$ means finite contribution at $s=-1$.
Thus, the studied divergence is exhibited through a pole of one of
the gamma functions in the Euler beta function of \req{limsumJm}.
If it had not been for the presence of the $s$ parameter,
\begin{enumerate}
\item the $r$ integral for the $q=2$ term would be logarithmically
divergent
(thus, each $\mathcal{J}_m^{(0)}$ would be infinite as $x_0\to\infty$),
\item the $m$ summation would not have been so easily reinterpreted as a
Riemann zeta function.
\end{enumerate}
To $\sum_{m=1}^{\infty} \mathcal{J}_m^{(0)}$ we have to add its $m=0$
counterpart, which is $1/2 \, J_0^{(0)}$.
In view of the first line of \req{defJ0sI0}, and taking into
consideration \req{intrs},
\beq
\frac{1}{2} \, J_0^{(0)}
=\frac{1}{4} \, \int_{r_a}^{\infty} dr \, r^{-2-s} \,
\int_{0}^{\pi/2} d\varphi \, \xi^2(r, \varphi) .
\eeq
The angular integral is given
by the first formula in \req{intphis} setting $m=1$.
We take the $x_0\to\infty$ limit of that result and get
\beq
\int_{0}^{\pi/2} d\varphi \, \xi^2(r, \varphi) \longrightarrow
\xi_0^2 \, \frac{\pi}{2}
\eeq
(the same as in \req{limiphi}). Hence,
\beq
\frac{1}{2} \, J_0^{(0)} \longrightarrow \xi_0^2 \, \frac{\pi}{8} \,
\int_{r_a}^{\infty} dr \, r^{-2-s}
= \xi_0^2 \, \frac{\pi}{8} \, \frac{r_a^{-(s+1)}}{s+1} ,
\label{limJ0}
\eeq
i.e., $J_0^{(0)}$ is singular for $x_0\to\infty$ and, like in the
previous case, this singularity translates into a pole at $s=-1$.
Finally, the sum of \req{limsumJm} and \req{limJ0} yields
\beq
\begin{array}{rrl}
\ds \sum_{m=1}^{\infty} \mathcal{J}_m^{(0)}+\frac{1}{2} \,
J_0^{(0)}&\longrightarrow&\ds \xi_0^2 \, \frac{\pi}{8} \left[
\zeta_R(s+1) \, \B\left( \frac{3-s}{2}, \frac{s+1}{2} \right)
+\frac{r_a^{-(s+1)}}{s+1}
+\mathcal{O}((s+1)^0)
\right] \\
&=&\ds\xi_0^2 \, \frac{\pi}{8} \left[
\frac{-1+r_a^{-(s+1)}}{s+1}
+\mathcal{O}((s+1)^0)
\right] \\
&=&\ds \mathcal{O}((s+1)^0) .
\end{array}
\label{sumfin}
\eeq
In other words, for $x_0\to\infty$ everything is finite at $s=-1$\footnote{
Although formula \req{sumfin} may give the impression that the first
finite contribution to $F$ is of the order of $\xi_0^2$, there are other
finite parts. Including them, and taking into account
the numerical differences between exact and approximate forms,
all terms proportional to $\xi_0^2$ cancel out ---within some
accuracy--- leaving just $\mathcal{O}(\xi_0^4)$, in
agreement with refs.\cite{MNN,KR}.}.
The second line in \req{sumfin} has been reached using
$\zeta_R(0)=-1/2$ and the
Laurent expansion for the divergent gamma function present in the beta
function. Note that the final line is independent of specific value
of $r_a$, provided that $r_a > 0$ ($r_a$ was the $r$ value from which
the exact integrand was replaced with its asymptotic form). This feature
is welcome, because it suggests that the shown finiteness does not
depend on minor details of the regularization procedure.

The finite character of \req{sumfin} is the consequence of cancelling
two singularities: one from the $\sum_{m=1}^{\infty}$ summation and the
other from the $m=0$ term.  Therefore, the price paid is that one loses
sight of the $m$ contributions as individually defined
entities.

\end{document}